\documentclass[a4paper,12pt]{article}
\usepackage[english]{babel}
\usepackage[final]{epsfig}
\usepackage{cite,subfigure}
\date{}
\begin{document}
\title{
\Large{
\textbf{Incorporation of a non-amphiphilic nematic
liquid crystal into a host monolayer}
}
}
\author{
Valentina S. U. Fazio, L. Komitov, S. T. Lagerwall\\
\textit{\small Liquid Crystal Physics, MINA, Chalmers University of
Technology}\\ 
\textit{\small \& G\"oteborg University, S-41296 G\"oteborg, Sweden}
\normalsize\\
\,\,\\
D. M\"obius\\
\textit{\small Max-Plank Institute for Biophysical Chemistry, P.O.Box
2841}\\
\textit{\small D-3400 G\"ottingen, Germany}
}
\maketitle
\normalsize
\begin{abstract}
Many nematic liquid crystals are not able to form stable monolayers at the
air/water interface because of the lack of a polar headgroup.
A possible way to obtain a monomolecular film with these compounds is to
incorporate them into host monolayers of amphiphilic compounds.
Stable monolayers containing a high fraction of the liquid crystal can be
obtained. 
We have prepared stable and transferable monolayers of MBBA (which is not an
amphiphilic compound) using octadecylmalonic acid (OMA) as host.
The monolayers at the air/water interface 
have been characterized by measurements of surface-pressure/area and 
surface-potential/area isotherms.
The monolayers deposited on quartz plates have been characterized by 
determining the transfer ratio and by spectroscopic measurements.
\end{abstract}

\vspace{5mm}

\section{Introduction}

\noindent
Organized monolayers at the air/water interface are usually formed by 
spreading amphiphilic molecules, i.e. molecules that are composed of a 
hydrophilic head group and a hydrophobic tail 
\cite{Gaines, Petty, BucDreFleKuh67, KuhnMobiusBucher, KuhnMobius}.
Such monolayers can be transferred sequentially to solid substrates 
for the construction of designed functional assemblies \cite{KuhnMobius}
or to prepare multilayer systems (Langmuir-Blodgett films).
The lack of a hydrophilic head group normally disqualifies a molecule 
for the formation of organized monolayers at the air/water interface.
However, in some cases such molecules as quinthiophene have been 
incorporated in a matrix monolayer of amphiphilic molecules 
\cite{SchTewKuh74}.
Another possible way to obtain monomolecular films of hydrophobic 
non-amphiphilic compounds is to incorporate them into host monolayers 
providing appropriate cavities\cite{VogMob88a, VogMob88b, CorMob92}.

The nematic liquid crystal MBBA 
($N$-(4-methoxybenzylidene)-4-butyla-
niline) 
is not amphiphilic and no 
ordered layers of very well defined thickness of this compound can 
be deposited using the Langmuir-Blodgett (LB) technique.
Nevertheless, we have been able to prepare stable two-component monolayers 
with a high fraction of MBBA incorporating it into host monolayers of the 
bifurcated amphiphile octadecylmalonic acid (OMA).
These two-component monolayers have been transferred onto solid 
substrates and the organization of the MBBA molecules in the 
monolayers has been investigated by absorption spectroscopy.

\section{Experiment}

\noindent
Monolayers at the air/water interface were prepared with a KSV3000 
Lang-
muir-Blodgett trough held in a cleanroom environment.
Ultrapure Milli-Q water (ph 5.5) was used.
The molecular arrangement of the monolayers at the air/water interface 
has been studied by measuring the surface-pressure/area 
(Wilhelmy method) and surface-potential/area (vibrating capacitor 
method) isotherms. 

LB monolayers were deposited onto quartz glass plates.
The transfer ratio, i.e. the ratio of monolayer area removed from the 
water and solid area coated, was monitored during the deposition and the 
LB films were characterized by measuring absorption spectra with a 
spectrophotometer.

\section{Results and discussion}

\subsection{Monolayers at the air/water interface}

\begin{figure}
\begin{center}
\epsfig{file=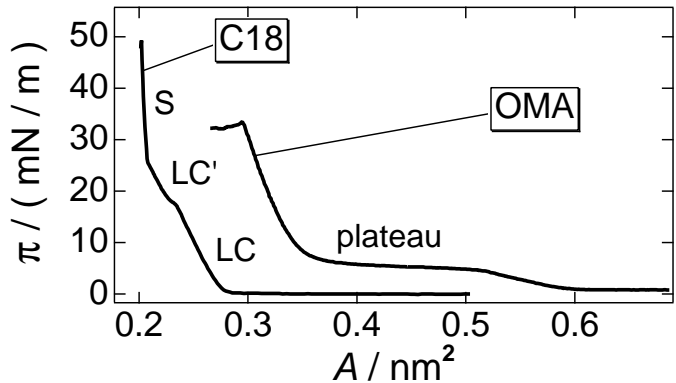, width=0.75\textwidth}
\caption{\small
\label{C18-OMA}
Comparison between the 
surface-pressure/area ($\pi / A$) isotherms of C18 and OMA.
}
\end{center}
\end{figure}

In Figure \ref{C18-OMA} the surface-pressure/area ($\pi/A$)
isotherm of OMA is compared with the well known isotherm 
of stearic acid (C18).
The two isotherms are quite different.
In the case of C18 the packing of the molecules, and thus
shape of the isotherm, is mainly due to the interactions between
the hydrocarbon chains.
Two liquid-condensed phases (\textsf{LC} and \textsf{LC}$^{\prime}$) 
which differ in the orientation of the chains, and a more condensed 
phase, also called solid phase (\textsf{S})  in which the carboxylic 
chains are vertically oriented, can be identified\cite{Gaines, Petty}.
Collapse takes place at a surface-pressure of 50\,mN/m.
On the other hands, in the case of OMA,
as pointed out by V. Vogel \textit{et al.}\cite{VogMob88a, VogMob88b}, 
because of the presence of two carboxylic acid groups in the same molecule,
the shape of the isotherm is determined by the polar headgroups.
The isotherm shows a plateau between 0.51 and 0.38\,nm$^{2}$/molecule.
Collapse takes place at a surface-pressure of 33\,mN/m.

\begin{figure}[t]
\begin{center}
\begin{minipage}{\textwidth}
\parbox[]{0.48\textwidth}{
\subfigure[]{
\epsfig{file=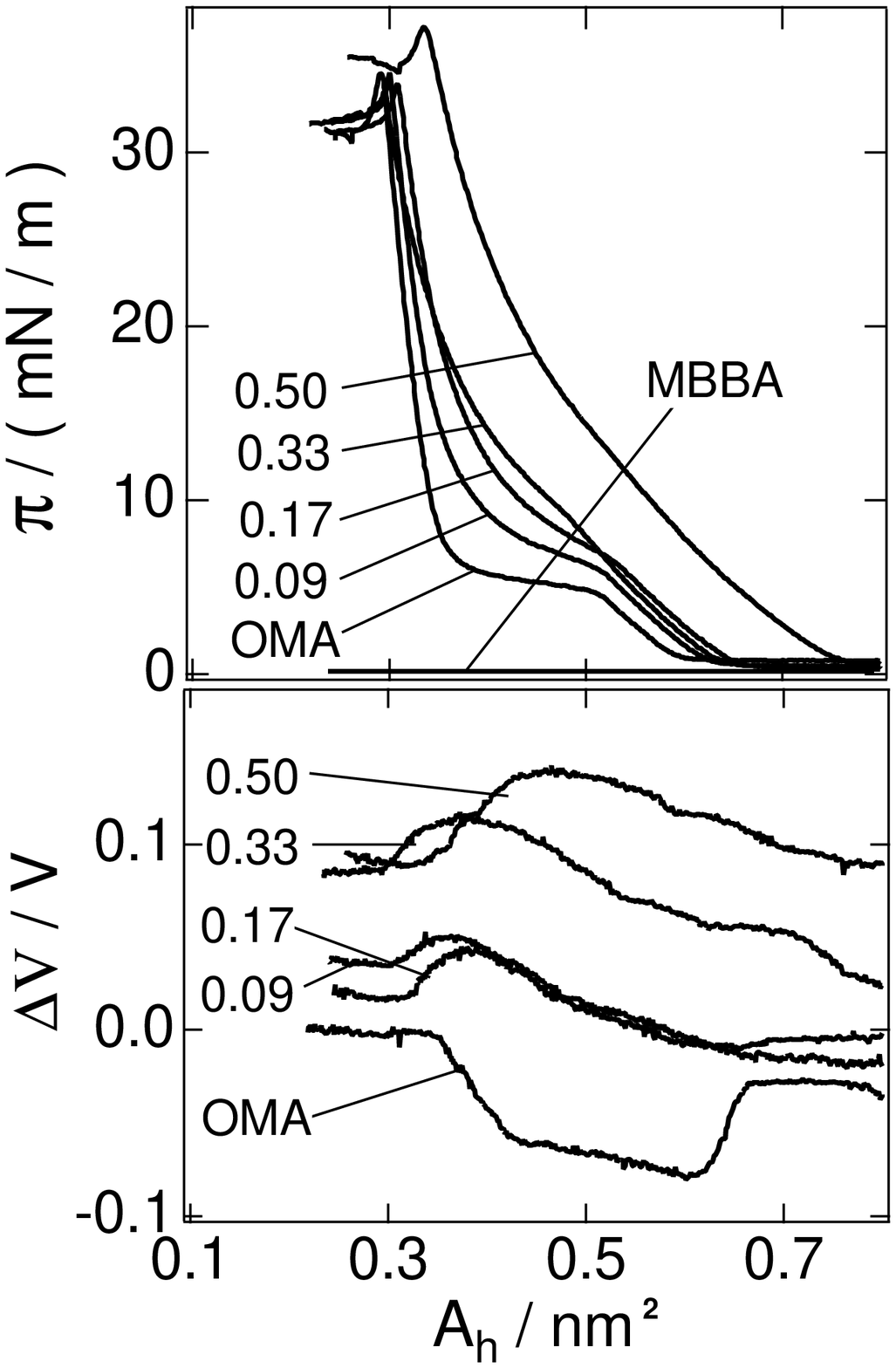, width=0.47\textwidth}
}
}
\parbox[]{0.51\textwidth}{
%\subfigure[]{
%\epsfig{file=collapse.eps, width=0.48\textwidth}\\
\subfigure[]{
\epsfig{file=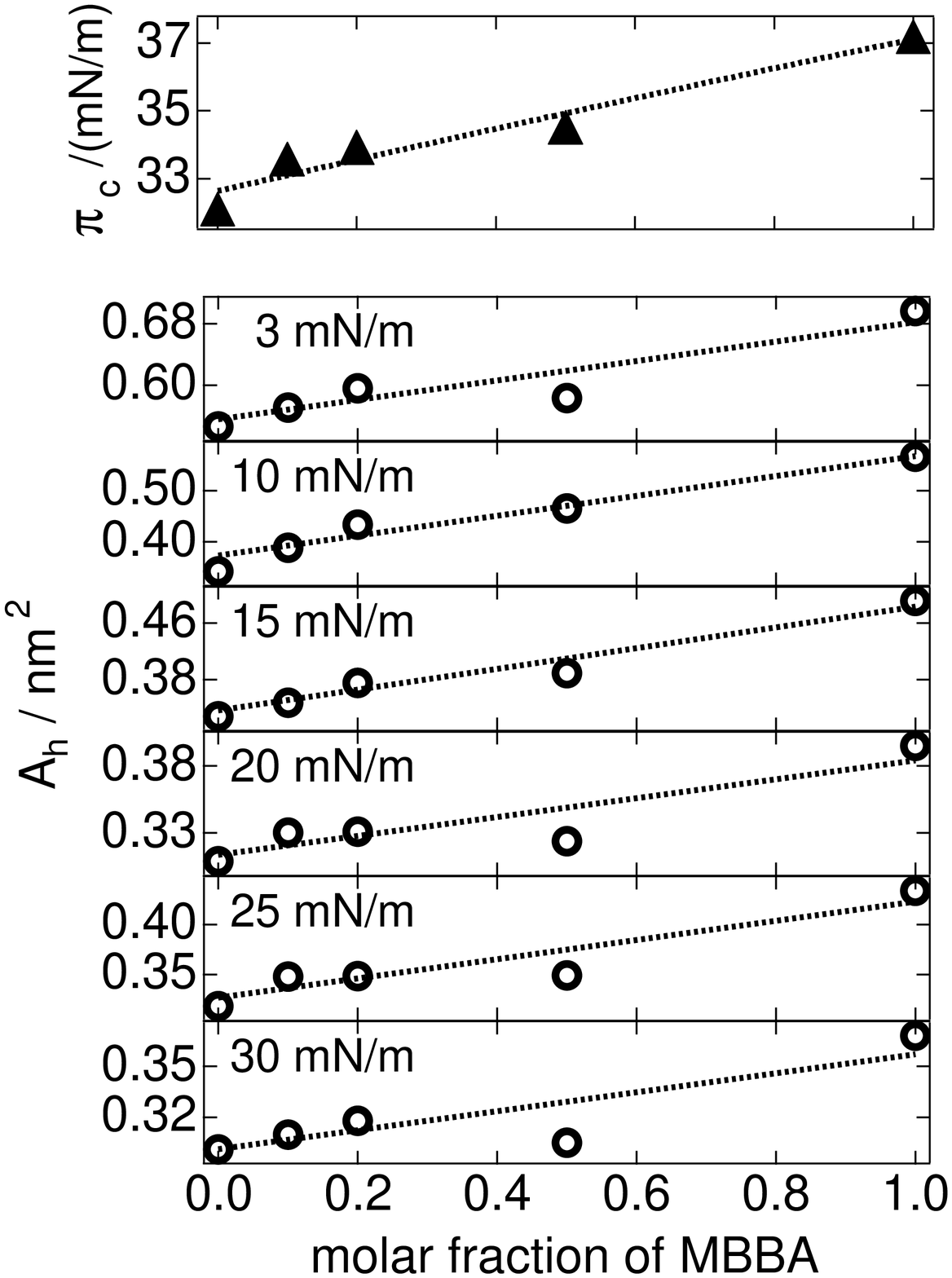, width=0.48\textwidth}
}
}
\end{minipage}
\caption{\small
\label{OMA-MBBA}
(a) Surface-pressure/area-per-host-molecule ($\pi / A_{h}$) and 
surface-potential/area-per-host-molecule ($\Delta V / A_{h}$)
isotherms for different molar fractions of MBBA, and for pure OMA and pure
MBBA.
(b) Dependence of the collapse pressure and the area-per-host-molecule on
the molar fraction of MBBA.
}
\end{center}
\end{figure}

MBBA is a non-amphiphilic compound and thus cannot form stable 
monolayers at the air/water interface (the surface-pressure/area 
isotherm of pure MBBA is shown in Figure \ref{OMA-MBBA}(a)).
We prepared mixed solutions of OMA and MBBA with different molar 
fractions of MBBA and studied the behavior of the mixed monolayers 
at the air/water interface.
The surface-pressure/area-per-host-{mo}{le}{cu}{le} ({$\pi / A_{h}$}) and 
surface-potential/area-per-host-molecule ({$\Delta V / A_{h}$}) isotherms 
for the mixtures are shown in Figure \ref{OMA-MBBA}(a),
where the area-per-host-molecule, $A_{h}$, is defined as:
\begin{displaymath}
A_{h} = 
\frac{\mbox{areaof the trough}}{\mbox{number of host molecules}}.
\end{displaymath}
The addition of MBBA to the OMA matrix influences considerably 
OMA's {$\pi / A$} isotherm.
On increasing the molar fraction of MBBA the plateau shrinks, 
the collapse pressure increases, and the shape of the isotherm 
resembles more and more that of a fatty acid.
Moreover, all isotherms in Figure \ref{OMA-MBBA}(a) are stable, 
which means that the surface-pressure is maintained with time 
at all areas.
For molar fractions of MBBA larger than 0.5 MBBA
molecules in excess are squeezed out of the monolayer 
during compression and the isotherms are destabilized. 

\begin{figure}[t]
\begin{center}
\epsfig{file=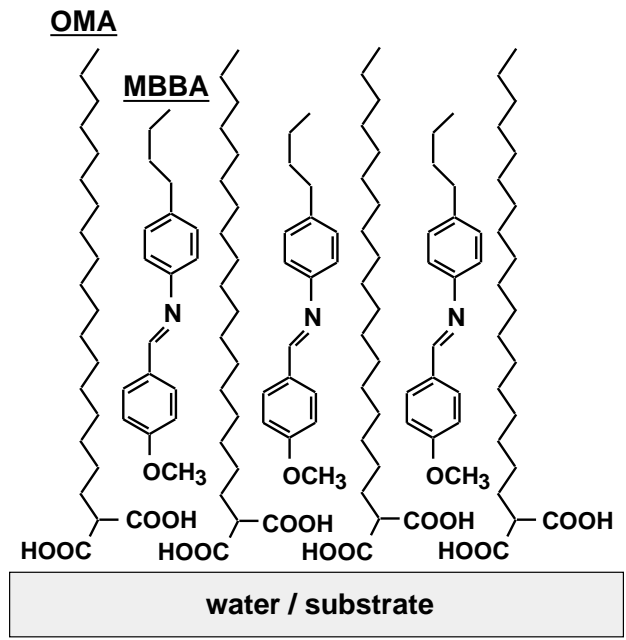, width=0.6\textwidth}
\caption{\small
\label{molecole}
Scheme of MBBA molecules incorporated into an OMA matrix.
}
\end{center}
\end{figure}

To distinguish between additive and cooperative incorporation of 
MBBA in OMA we have plotted the collapse pressures $\pi_{c}$, and the 
areas-per-host-molecule $A_{h}$ at different surface pressures, versus
the molar fraction of MBBA in Figure \ref{OMA-MBBA}(b).
If the incorporation is additive $\pi_{c}$ and $A_{h}$ of mixed 
OMA/MBBA monolayers should follow a linear relation with the 
molar fraction of one of the two substances\cite{Gaines, CorMob92}.
Any deviation from this behavior would be an evidence of 
cooperative incorporation.
In our case, as we can see from the figures, $\pi_{c}$ and $A_{h}$
follow an almost linear relation with the molar fraction of MBBA
which indicates that the incorporation is additive:
MBBA molecules penetrate into the chain region of the OMA 
monolayer as depicted in Figure \ref{molecole}.

The presence of MBBA also influences the $\Delta V/A_{h}$ isotherms.
In general\cite{Gaines, Petty, VogMob88a, VogMob88b} the surface potential
is proportional to the change of the normal component of the dipole density
with respect to the pure water.
For OMA $\Delta V$ is expected to be negative because this molecule has two
polar heads ($\mu_{COOH}= - 0.201$\,D) and one 
terminal CH$_{3}$ group  ($\mu_{CH_{3}}= + 0.351$\,D).
With the addition of MBBA only the number of CH$_{3}$ groups increases 
and, as a consequence, the surface potential also increases and becomes 
positive.

The dipole moment of a CH$_{3}$ group at the interface with air has the 
same sign as at the interface with water.
Thus, the $\Delta V/A_{h}$ isotherms do not tell us about the orientation 
of MBBA molecules in the OMA matrix.
Here we assume that MBBA molecules orient as in Figure \ref{molecole} since
the COH$_{3}$ groups are more hydrophilic than the CH$_{3}$ ones and 
thus it is reasonable to think that they prefer to orient toward the 
polar headgroups of the host molecules\cite{CorMob92}.

\subsection{Deposition}

We have deposited monolayers of pure OMA and of the {OMA:MBBA 1:1} 
mixture onto quartz glass plates.

OMA can be deposited at all surface pressures with unitary 
transfer ratio, provided that the deposition speed is not too large 
(not more than 5-6\,mm/min).
On the other hands, OMA:MBBA\,1:1 monolayers resulted more 
difficult to be deposited without molecular rearrangement or collapse
during the transfer onto the substrates.
Deposition pressure, deposition speed, and substrate characteristics 
(we have also performed depositions onto ITO coated glass plates 
that are not reported here) can influence the transfer ratio
as shown in Table \ref{dep_table}. 
\begin{table}
\begin{center}
\caption{\small
Transfer ratios of the OMA:MBBA\,1:1 monolayers on quartz 
for different deposition pressures and deposition speeds.
For a prefect deposition the transfer ratio should be 1.
\label{dep_table}
}
\smallskip
\begin{tabular}{|c|c|c|}
\hline
deposition & deposition & transfer \\
 pressure & speed & ratio \\
\hline
\hline
10\,mN/m & 10\,mm/min & \textbf{1}\\
\hline
10\,mN/m & 5\,mm/min & \textbf{1} \\
\hline
15\,mN/m & 10\,mm/min & \textbf{1} \\
\hline
15\,mN/m & 5\,mm/min & \textbf{1} \\
\hline
20\,mN/m & 10\,mm/min & $\ll$1 \\
\hline
20\,mN/m & 5\,mm/min & $\ll$1 \\
\hline
20\,mN/m & 2\,mm/min & $\ll$1 \\
\hline
\end{tabular}
\end{center}
\end{table}
Good transfer ratios were obtained on quartz when depositing the
mixed OMA:MBBA\,1:1 monolayers at 10 and 15\,mN/m at all deposition speeds. 

\begin{figure}
\begin{center}
\epsfig{file=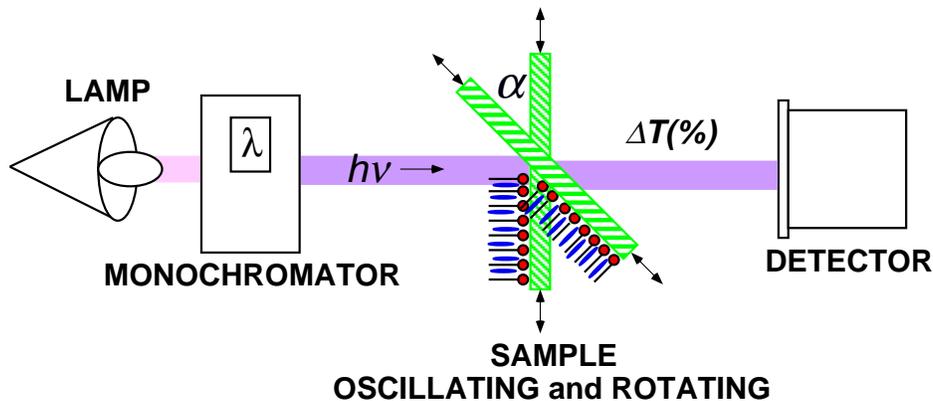, width=0.9\textwidth}
\caption{\small
\label{setup_spectra}
Setup for measuring absorption spectra of LB monolayers.
The light coming from a UV-lamp is first sent through a monochromator and 
and then to the sample.
The sample oscillates so that the light beam hits alternatively the clean
part of the glass plate and the part coated with the LB monolayer.
The difference in light transmission between the two cases ($\Delta T$) is
collected by a detector. 
The sample can also be rotated to perform the analysis at different 
incidence angles $\alpha$.
}
\end{center}
\end{figure}

\begin{figure}[t]
\begin{center}
\subfigure[]{\epsfig{file=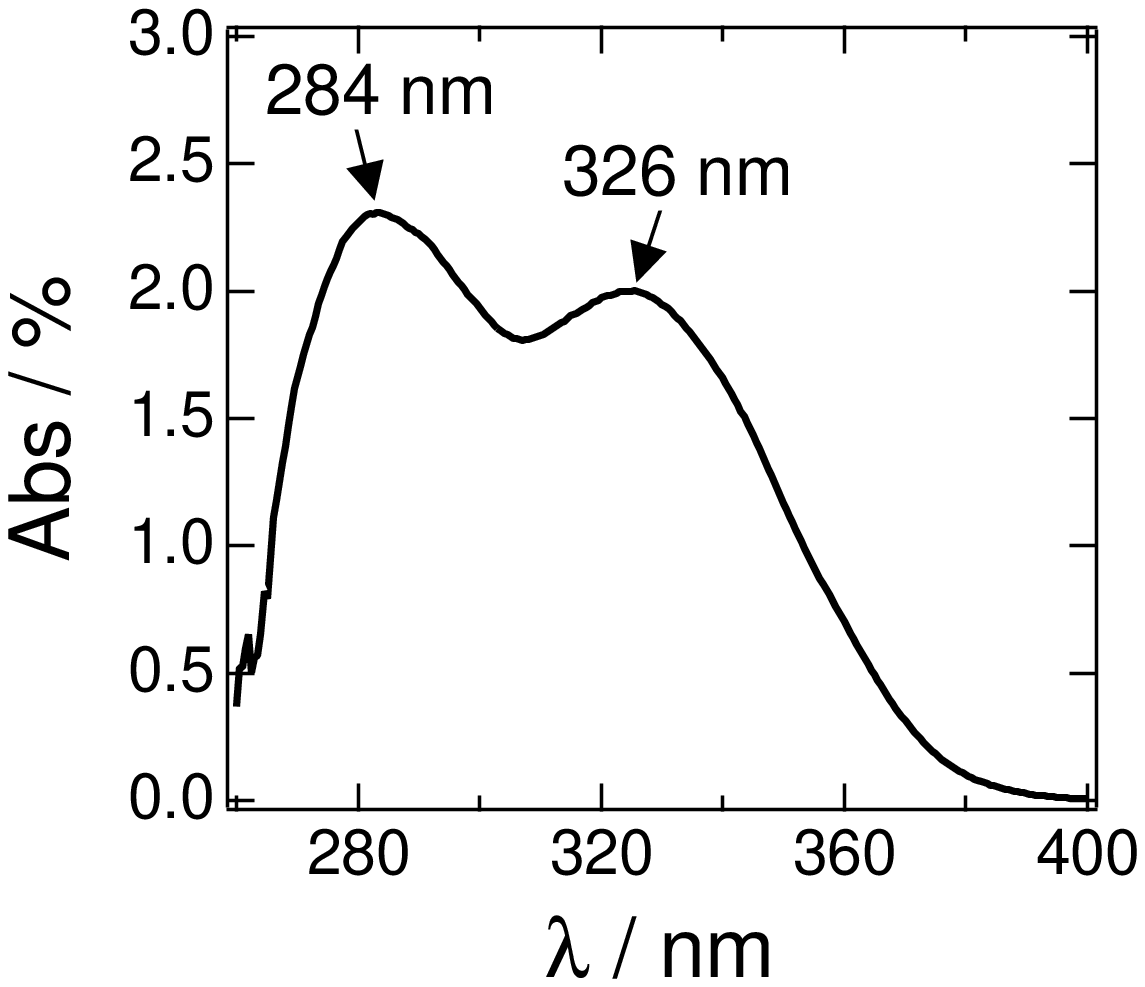, width=0.38\textwidth}}
\hfill
\subfigure[]{\epsfig{file=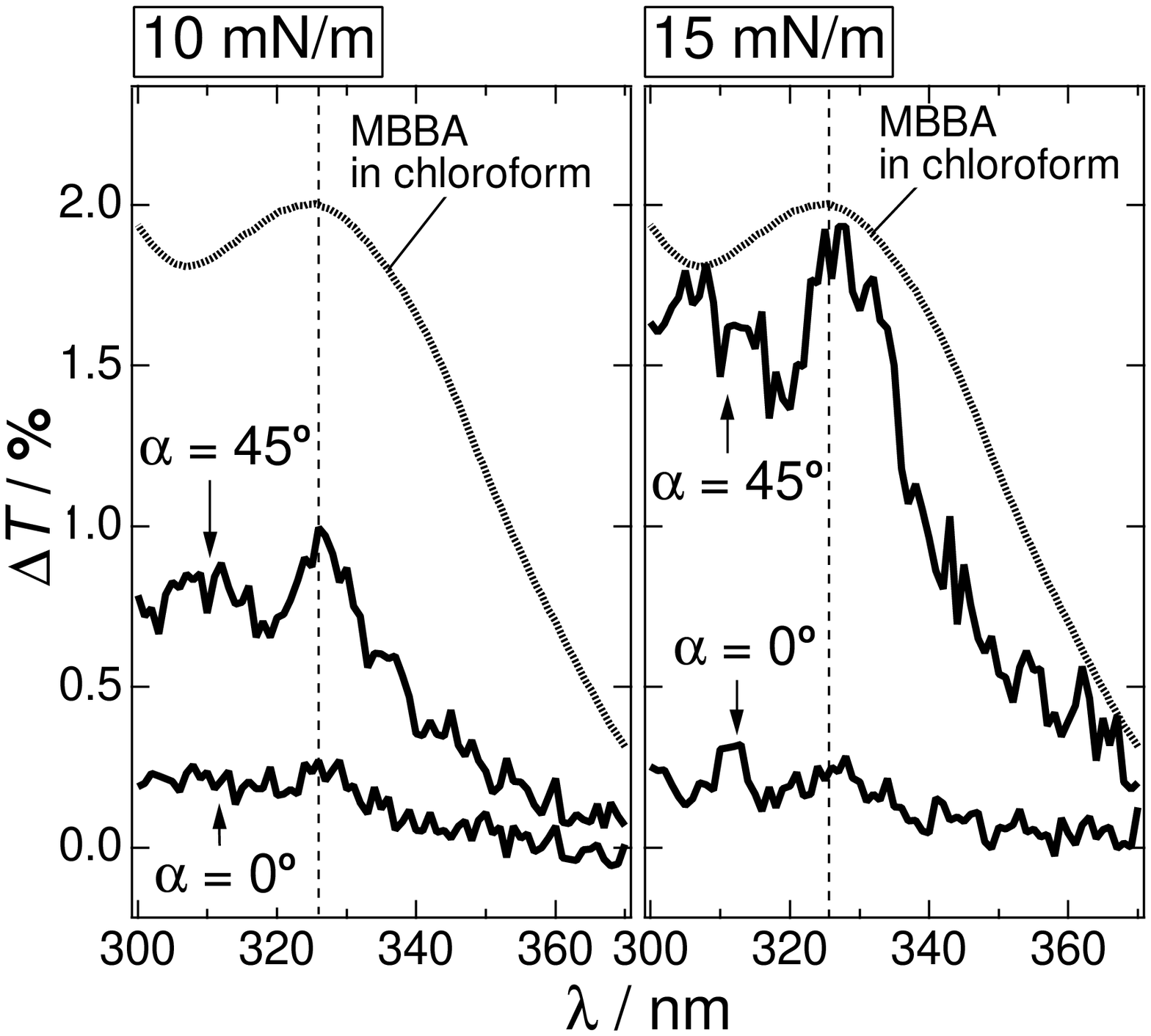, width=0.60\textwidth}}
\caption{\small
\label{OMA_MBBA_spectra}
(a) Absorption spectrum of MBBA in chloroform (0.13\,mM). The spectrum
presents two peaks, at 284\,nm and at 326\,nm.
(b) Spectra of OMA:MBBA\,1:1 LB monolayers on quartz glass plates deposited
at two different surface-pressures. 
$\alpha$ is the angle of incidence of the light as depicted in \protect{Figure
\ref{setup_spectra}}. 
At normal incidence ($\alpha = 0^{\circ}$) there is almost no absorption,
while a structure is present for $\alpha = 45^{\circ}$.
}
\end{center}
\end{figure}

However, transfer ratio measurements are not sufficient for the
characterization of the films.
We recorded UV spectra of the monolayers with the setup depicted in Figure
\ref{setup_spectra}. 
The spectra are shown in Figure \ref{OMA_MBBA_spectra}(b) for mixed 1:1 
monolayers deposited at two different surface pressures and at the 
deposition speed of 10\,mm/min (first and third rows in Table 
\ref{dep_table}).

At normal incidence ($\alpha = 0^{\circ}$ in Figure \ref{setup_spectra})
only a very small fraction of the 
light is absorbed by the mixed LB film and the spectra are almost flat.
At 45 degrees angle of incidence instead, and for the same 
monolayers, the spectra present a structure and a peak at 326\,nm.
This tells us not only that MBBA is present in the LB film, but also 
that it is homeotropically aligned.
Indeed, at normal incidence the polarization of the light lies in the plane of 
the glass plate perpendicular to the long axis of the MBBA molecules, and
light is not absorbed.
At any other incidence angle the polarization of the light has a 
component along the long molecular axes which is absorbed.

The fact that the films deposited at the higher surface-pressure 
absorb more can be explained in terms of molecular density.
In films deposited at 15\,mN/m surface-pressure 
the molecules are more densely 
packed and the number of MBBA molecules per unit area is about 1.6 
times that of films deposited at 10\,mN/m.
The first films are thus expected to show absorption spectra
about 60\,{\%} more intense. 
This explains, in the limits of the experimental error\cite{note1}, the 
differences in the spectra of Figure \ref{OMA_MBBA_spectra}(b).

\section{Conclusions}

\noindent
We have deposited Langmuir-Blodgett monolayers of the non-amphiphilic nematic 
liquid crystal MBBA incorporating it into a host monolayer of 
octadecylmalonic acid (OMA).
By studying the molecular arrangement at the air/water interface 
MBBA resulted to be additively incorporated into the host 
monolayer.
Mixed monolayers OMA:MBBA\,1:1 could be deposited at certain 
surface-pressures and deposition speeds without molecular 
rearrangement and/or collapse.

\section{Acknowledgments}

The authors would like to thank dr. H. Huesmann for assistance in the
experimental work.
Valentina S. U. Fazio would like to thank the TMR programme (contract number 
ERBFMBICT983023) for financial support.

\bibliography{journal2,moeb}

\begin{thebibliography}{10}
\bibitem{note1}
The experimental error is here mainly due to the fact that we compare two LB
  films that result from different depositions.
\end{thebibliography}

\end{document}